\documentstyle[12pt,a4,fleqn]{article}

\begin{document}
\date{}
\title{Aharonov-Bohm interference in the presence of\\
 metallic mesoscopic cylinders}
\author{Silviu Olariu\\
Institute of Physics and Nuclear Engineering, Sectia a III-a\\
76900 Magurele, Casuta Postala MG-6, Bucharest, Romania}
\maketitle
\abstract
This work studies the interference of electrons in the presence of a line of 
magnetic flux surrounded by a normal-conducting mesoscopic cylinder at low 
temperature. It is found that, while there is a supplementary phase 
contribution from each electron of the mesoscopic cylinder, the sum of these 
individual supplementary phases is equal to zero, so that the presence of a 
normal-conducting mesoscopic ring at low temperature does not change the 
Aharonov-Bohm interference pattern of the incident electron. It is shown that
it is not possible to ascertain by experimental observation that the 
shielding electrons have responded to the field of an incident electron,
and at the same time to preserve the interference pattern of the incident
electron. It is also shown that the measuring of the transient magnetic field
in the region between the two paths of an electron interference experiment
with an accuracy at least equal to the magnetic field of the incident
electron generates a phase uncertainty which destroys the interference
pattern.
\endabstract

\section{Introduction}
In an Aharonov-Bohm (AB) experiment, \cite{1} the incident electrons are 
prevented from entering the region of the magnetic flux by certain shields.
If these shields are superconducting cylinders or mesoscopic cylinders at
low temperature, the shielding electrons occupy states which possess phase
coherence around the line of magnetic flux. These shielding electrons could
in principle bring an additional phase contribution to the conventional
AB phase shift of the incident electrons. Experiments carried out by Lischke,
\cite{2} M\"{o}llenstedt et al. \cite{3} and Tonomura et al. \cite{4} have
demonstrated however the persistence of the conventional AB shift in the
presence of metallic shields.\\

The absence of a supplementary phase shift due to the shielding electrons has
been explained in the case of a metallic shield at normal temperature by
Peshkin, \cite{5} who pointed out that one could imagine the conductor as being
cut, so that the vector potential of the enclosed flux which acts on the
{\it shielding} electrons can be gauged away. Moreover, Goldhaber and
Kivelson \cite{6} \cite{7} have shown that there are no additional phase shifts
due to the electrons of a supeconducting shield, because of the $2e$ charge
of the electron pairs and of the quantization of the magnetic flux.\\

The case when the shield is a metallic mesoscopic cylinder at low temperature
is studied in the present work. In a mesoscopic ring, the phase coherence
length may be comparable or larger than the circumference of the ring, as
demonstrated by the oscillations of the magnetoconductance, with a magnetic
flux period $h/2e$. \cite{8}-\cite{11} It is found in this work that, although
there is a supplementary phase contribution from each electron of the 
mesoscopic cylinder, the sum of these individuals supplementary phases is equal
to zero, so that the presence of a normal-conducting mesoscopic cylinder at low
temperature does not change the AB interference pattern of the incident 
electron. \\

The shields in the AB experiments are supposed to prevent the overlap between
the incident electrons and the line of magnetic flux {\it and} to screen the
electromagnetic fields of the incident electrons. It is shown that
it is not possible to ascertain by experimental observation that the 
shielding electrons have responded to the field of an incident electron,
and at the same time to preserve the interference pattern of the incident
electron. It is also shown that the measuring of the transient magnetic field
in the region between the two paths of an electron interference experiment
with an accuracy at least equal to the magnetic field of the incident
electron generates a phase uncertainty which destroys the interference
pattern.\\

The discussion in Sec. 2 of the classical interaction of an incident electron
with a charged rotator and a line of magnetic flux serves as basis for the
determination in Sec. 3 of the supplementary quantum-mechanical phase shift
due to the interaction of an electron with the line of magnetic flux and 
in the presence of the charged rotator. The interaction of an incident electron
with a line of magnetic flux in the presence of a circular metallic string
at 0 K is discussed in Sec. 4. In Sec. 5 it is shown that the supplementary
phase shift for a metallic circular string decreases exponentially with 
temperature above 0 K. In Sec. 6 it is found that  the supplementary 
contribution averages to zero in the case of a metallic mesoscopic hollow
cylinder of non-zero thickness and height. The limitations inherent to the
process of observation of weak transient magnetic fields are discussed in
Sec. 7.\\

\section{Classical interaction of an incident electron with 
a line of magnetic flux and a charged rotator}

The interaction of an incident electron with a line of magnetic flux in the
presence of a mesoscopic ring is schematically represented in Fig. 1. The 
electric field of an incident electron exerts an action on the electrons of the
mesoscopic cylinder, so that the motion of these shielding electrons is
correlated, or coherent with the motion of the incident electron on one or the
other side of the line of enclosed flux. Supplementary flux-dependent phase
shifts could then be expected in principle from these shielding electrons,
in addition to the conventional AB phase shift due to the interaction of the
incident electron with the line of magnetic flux. The simpler case which will
be analyzed in this section is the classical interaction of an incident
particle of charge $q$ and velocity $v>0$ which moves along a straight line
passing at a distance $d$ from the axis of the enclosed magnetic flux $F$,
while  a charge $Q$ uniformly distributed and rigidly attached to a ring of
radius $R$ can freely rotate with angular velocity $\Omega$ round the axis
of the magnetic flux, as shown in Fig. 2. For $d \gg R$, the Lagrange function
of this system is, in SI units,
\begin{equation}
L(v,\Omega)=\frac{1}{2}m v^2+\frac{1}{2} M R^2 \Omega^2+
\frac {qQ R^{2} d \Omega v}{8 \pi \epsilon_0 c^{2} 
\left( x^{2}+d^{2} \right)^{3/2}}+ 
\frac {qFdv}{2 \pi \left( x^{2}+d^{2} \right)} + \frac{1}{2\pi} QF \Omega ,
\end{equation}
where $m$ is the mass of the incident particle and $M$ the mass of the 
charged rotator. The last three terms in Eq. (1) represent respectively the
interaction between the charged ring and the incident particle; between the 
magnetic flux and the incident particle; and between the magnetic flux and
the charged ring. These contributions are obtained as the product of a charge,
of a vector potential and of a velocity, the vector potential of the charged
ring being calculated in the magnetic dipole approxination, valid as long as
$d \gg R$. It can be shown from the Lagrange equations that
\begin{equation}
\Omega + \frac{a(x) v}{M R^2} = \Omega_0 ,
\end{equation}
\begin{equation}
v^2 \left[ 1 - \frac{a^2 (x)}{mM R^2} \right] = v_0^2,
\end{equation}
where $a=a(x)$ is given by
\begin{equation}
a(x) = \frac{qQ R^2 d}{8 \pi \epsilon_0 c^2 (x^2+d^2)^{3/2}} ,
\end{equation}
and where $\Omega_0$ and $v_0$ are respectively the angular velocity of the
charged ring and the velocity of the incident particle when the distance 
between the ring and the particle is very large. If $q$ and $Q$ are equal to
the electron charge $-e<0$ and the masses $m,M$ are equal to the electron mass,
it results from Eqs. (2)-(4) that the parameter $a(x)$ is proportional to the
classical electron radius $r_0=2.8\cdot 10^{-15}$ m, so that the variation
$\Omega-\Omega_0$ of the angular velocity of the ring is proportional to
$r_0$, while the variation $v-v_0$ of the velocity of the incident electron
is proportional to $r_0^2$. Therefore, in a first-order approximation with
respect to $r_0$, we can consider that the velocity of the incident electron
is constant. The enclosed magnetic flux $F$ does not appear in Eqs. (2)-(4),
which means that there are no observable classical effects of an enclosed
magnetic flux.\\

\section{Supplementary flux-dependent phase shift in the
presence of a charged rotator}

In Fig. 3 is represented a system composed of an incident particle of charge
$q$ moving along a straight line, of a tube of magnetic flux $F$, and of a
particle of charge $Q$ and mass $M$ which can move on a circle of radius $R$.
The non-relativistic, quantum-mechanical evolution of this system will be
analyzed by assuming that the incident particle is moving with constant
velocity $v$ along its straight path. The phase shift will be determined by
considering that the magnetic field of the incident particle and the vector
potential of the enclosed flux are given functions of space and time, and 
shall neglect the irrotational component of the electric field of the incident
particle.\\

The Schr\"{o}dinger equation for the wave function $\Psi(\theta,t)$ of the
particle of charge $Q$ is, in a first-order approximation with respect to
the parameter $qQ/(4\pi \epsilon_0 Mc^2)$,
\begin{equation}
i\hbar \frac{\partial \Psi}{\partial t} = \frac{1}{2MR^2}\left(-i\hbar
\frac{\partial}{\partial \theta} - \frac{QF}{2\pi}\right)^2 -
\frac{qQdv}{8\pi\epsilon_0 Mc^2 \left(v^2t^2+d^2\right)^{3/2}}\left(
-i\hbar\frac{\partial}{\partial\theta} - \frac{QF}{2\pi}\right) .
\end{equation}
The quantity  $(-i\hbar \partial/\partial \theta - QF/2\pi)/MR^2$ is the
operator for the angular velocity of the particle of charge $Q$, and the last
term in Eq. (5) corresponds to the energy of interaction between the magnetic
moment of the rotating charge $Q$ and the magnetic field of the incident
particle, written as a function of time. The solutions of Eq. (5) are of the
form
\begin{equation}
\Psi_n(\theta,t)=e^{in\theta-i\Phi_n(t)} ,
\end{equation}
where the phase $\Phi_n(t)$ is given by
\begin{equation}
\Phi_n(t)=\frac{\hbar}{2MR^2}\left(n-\frac{QF}{2\pi\hbar}\right)^2 t -
\frac{qQ(n-QF/2\pi\hbar)}{8\pi\epsilon_0Mc^2d}\frac{vt}{(v^2t^2+d^2)^{1/2}} ,
\end{equation}
$n$ being an integer. Thus, as the particle of charge $q$ passes from the 
incidence region ($t=-\infty$) to the observing region ($t=\infty$), Eq. (7)
shows that there is a supplementary phase shift $\delta_n$ given by
\begin{equation}
\delta_n=-\frac{qQ}{4\pi\epsilon_0Mc^2d}\left(n-\frac{QF}{2\pi\hbar}\right) .
\end{equation}

If the incident particle of charge $q$ is moving along a straight line situated
to the left of the enclosed flux, a path represented by the dashed line of
Fig. 3, then the supplementary phase shift due to the action of this particle 
on the charge $Q$, which is rotating in the presence of the enclosed flux $F$, 
will be $-\delta_n$. If both $q$ and $Q$ are electrons, then the interference
pattern of the incident electron will be shifted by
\begin{equation}
\Delta_n=\Delta_{AB}^{(e)}-\frac{2r_0}{d}\left(n-\frac{QF}{2\pi\hbar}\right),
\end{equation}
where $\Delta_{AB}^{(e)}=-eF/\hbar$ is the conventional AB shift for the 
incident electron. The shift in Eq. (9) is independent of the radius $R$ of 
the string. An alternative way to obtain this result would be to regard the 
incident electron as moving in the {\it applied} vector potential of the 
enclosed flux $F$ {\it and} of the charged rotator.\\

\section{Supplementary flux-dependent phase shift in the presence of a
circular metallic string at 0 K}

A circular chain of atoms having the property that each atom contributes
one electron which can move freely along the chain will be referred to as a
circular metallic string. The analysis in this section of the interaction
of an incident electron with a tube of magnetic flux in the presence of
the one-dimensional circular metallic string will be used in Sec. 6 to study
the AB interactions in the presence of real metallic rings.\\

The state of the electrons in the circular metallic string is described by
a multielectron antisymmetric wave function 
$\Psi_{\rm meso}(\theta_1,\theta_2,...\theta_{N_0},t)$ depending on the angular
variables $\theta_1, \theta_2, ...\theta_{N_0}$ which give the positions
of the $N_0$ electrons. The wave function $\Psi_{\rm meso}$ is a solution of 
the Schr\"{o}dinger equation
\begin{eqnarray}
\lefteqn{i\hbar\frac{\partial\Psi_{\rm meso}}{\partial t} = \frac{1}{2m_eR^2}
\sum_{j=1}^
{N_0}\left(-i\hbar\frac{\partial}{\partial\theta_j}+\frac{eF}{2\pi}\right)^2
\Psi_{\rm meso}}\nonumber\\
 & & -\frac{e^2dv}{8\pi\epsilon_0m_ec^2(v^2t^2+d^2)^{3/2}}
\sum_{j=1}^{N_0}\left(-i\hbar\frac{\partial}{\partial\theta_{j}}+
\frac{eF}{2\pi}\right)\Psi_{\rm meso} ,
\end{eqnarray} 
where the charge and the mass of the electron are $-e<0$ and $m_e$. The wave
function $\Psi_{\rm meso}$ can be written as a Slater determinant involving 
the single-electron states of angular momenta $m_1, m_2, ... m_{N_0}$. Since 
the single-electron states are the charged-rotator states of Sec. 3, the phase
shift in the interference pattern of the incident electron in the presence
of the circular metallic string can be obtained from Eq. (9) as
\begin{equation}
\Delta_N=\Delta_{AB}^{(e)}-\frac{2N_0r_0}{d}S_N ,
\end{equation}
where
\begin{equation}
S_N=\frac{1}{N_0}\sum_{j=1}^{N_0}\left(m_j+\frac{eF}{2\pi\hbar}\right) .
\end{equation}

At 0 K the single-electron states are occupied in the order of increasing
energy, and the states above the Fermi level are empty. The supplementary
phase shift $\Delta_N-\Delta_{AB}^{(e)}$ written in Eqs. (11) and (12) is a
{\it periodic} function of the enclosed magnetic flux $F$, the period being
$h/e$. If the number of electrons $N_0$ is odd and  $e|F|/2\pi\hbar
<1/2$, the occupied states have at 0 K the angular momentum numbers $0, \pm1,
\ldots \pm[N_0/2]$, where $[N_0/2]$ is the integer part of $N_0/2$, so that
\begin{equation}
S_N=eF/2\pi\hbar\,, \:\,{\rm for} \:\,N_0 \:\,{\rm odd}\:\, {\rm and}\:\,
  e|F|/2\pi\hbar<1/2\:.
\end{equation}
If $N_0$ is even and $0<eF/2\pi\hbar<1/2$, the occupied stated have the
quantum numbers $0,\pm1,...\pm(N_0/2-1), -N_0/2,$ so that 
\begin{equation}
S_N=-1/2+eF/2\pi\hbar \,,\:\,{\rm for} \:\,N_0 \:\,{\rm even}\:\, {\rm and}\:\,
  0<eF/2\pi\hbar<1/2\:,
\end{equation}
and if $N_0$ is even and $-1/2<eF/2\pi\hbar<0$, the occupied states have the
quantum numbers $0,\pm1,...\pm(N_0/2-1),N_0/2$, so that 
\begin{equation}
S_N=1/2+eF/2\pi\hbar \,,\:\,{\rm for} \:\,N_0 \:\,{\rm even}\:\, {\rm and}\:\,
  -1/2<eF/2\pi\hbar<0\:.
\end{equation}

The function $S_N$ is represented in Fig. 4 as a function of the magnetic flux
$F$ for a circular string at 0 K, assuming that the number of electrons 
appearing in Eq. (12) is very large. It can be seen from Fig. 4 that $S_N$,
and with it the phase shift in Eq. (11), is different for circular strings
containing an even number of electrons or an odd number of electrons.\\

The maximum value of the supplementary phase shift is, from Eq. (11),
$r_0/d$ per electron of the circular string. For $d$=1 $\mu$m, this maximum
value is $2.8\cdot 10^{-9}$ rad per electron.\\

\section{Temperature dependence of the supplementary phase shift for a
circular metallic string at 0 K}

The expression of the phase shift given in Eqs. (11) and (12) is valid provided
that the occupation number for the electron states changes abruptly from 1 
bellow the Fermi level to 0 above the Fermi level. For temperatures of the
metallic string above 0 K, the occupation numbers are given by the Fermi-Dirac
distribution, and the phase shift in the interference pattern of the incident
electron will be
\begin{equation}
\Delta_T=\Delta_{AB}^{(e)}-\frac{2r_0}{d}C_T ,
\end{equation}
where
\begin{equation}
C_T=\sum_{n=-\infty}^{\infty}\frac{n+\frac{eF}{2\pi\hbar}}
{\exp\left[\frac{\frac{\hbar^2}{2m_eR^2}\left(n+\frac{eF}{2\pi\hbar}\right)^2
-E_0}{kT}\right]+1} ,
\end{equation}
and $E_0$ is the energy of the Fermi level for the circular
string containing $N_0$ electrons. If the temperature of the string is such 
that
\begin{equation}
\frac{N_0\hbar^2}{4\pi^2km_eR^2}\ll T\ll \frac{N_0^2\hbar^2}{8\pi km_e R^2 } ,
\end{equation}
where $k$ is the Boltzmann constant,
the sum in Eq. (17) can be evaluated with the aid of the Poisson sum formula
\cite{12} as
\begin{equation}
C_T=-\frac{4\pi m_eR^2kT}{\hbar^2}
\sin (2eF/\hbar) \exp \left(-\frac{4\pi^2m_eR^2kT}{N_0\hbar^2}\right) .
\end{equation}
Due to the condition (18), the factor $C_T$ is exponentially small, and unlike
the function $S_N$, Eq. (12), $C_T$ is not proportional to the number of
electrons of the string. Thus, the supplementary phase shift due to the global
interaction of the incident electron, magnetic flux and metallic string
vanishes rapidly with increasing temperature of the string. The AB shift is,
of course, independent of temperature. If we consider that $N_0=2\pi R/a_0$,
where $a_0$ is the interatomic distance, then the lower and upper limits in
Eq. (18) are, for $a_0=2.5 \cdot 10^{-10}$ m and $R=10^{-6}$ m, $N_0\hbar^2/
(4\pi^2km_eR^2)=0.56$ K and $N_0^2\hbar^2/(8\pi km_eR^2)=2.2\cdot 10^4$ K.

\section{Cancellation of the supplementary phase shift for a real metallic 
mesoscopic cylinder}

In the case of a mesoscopic cylinder of non-zero thickness and height having
the axis along the z-direction there are, in addition to the angular momentum 
quantum number, two quantum numbers resulting from boundary conditions on the
motion in the radial and the z-directions. For each pair of these radial and
z quantum numbers, the number of substates depending on the angular momentum
quantum number may be even or odd, as discussed in Sec. 4. The phase shift
in the interference pattern of the incident electron will be
\begin{equation}
\Delta_N=\Delta_{AB}^{(e)}-\frac{2Nr_0}{d}S_N ,
\end{equation}
where $N$ is the number of free electrons of the mesoscopic ring, and $S$ is 
the average of $S_N$ with the weights  $e|F|/\pi\hbar$ for an even number
of substates and $(1-e|F|/\pi\hbar)$ for an odd number of substates,
where we have assumed that $e|F|/\pi\hbar<1/2$. The weight $e|F|/\pi\hbar$ is
obtained as the ratio of the energy separation between the two sublevels
with the same $m$, which is proportional to 2$em|F|/\pi\hbar$, and the energy
separation between the substates of magnetic quantum numbers $m-1$ and $m$, 
this separation being proportional to $2m$. These weights and Eqs. (13)-(15)
then give
\begin{equation}
S=0\:,
\end{equation}
which means that, due to averaging, the AB interference pattern of the incident
electron is not changed by the presence of the metallic mesoscopic cylinder.\\

\section{Measurability of weak transient magnetic fields in electron
interference experiments}

The primary function of a shield in an AB experiment is of separating the
region of space accessible to the incident electrons from the region of the
magnetic flux. At the same time, a metallic cylinder of sufficient thickness
is expected to prevent the electric and magnetic fields of an incident electron
from entering the region of the magnetic flux. In a classical picture, this
screening action is explained by the motion of the electrons of the metallic
shield produced by the transient electric field of the incident electron.\\

The part of the electric field of the incident electron having a non-zero
circulation is of the order of $ev^2/(4\pi \epsilon_0 c^2 d^2)$, for a
velocity $v$ and at a distance $d$. If the distance between the
incident electron and a shielding electron is of the order of $d$, and the
velocity of the incident electron is $v$, the displacement of the shielding
electron under the action of the afore-mentioned part of the incident electric
field is, for a time $2d/v$, of the order of $2r_0$, the classical electron
radius, thus being extremely small.\\

If $N$ electrons are taking place to the shielding process, then in order to
be able to ascertain that they have indeed responded to the field of the
incident electron, their average position must be known with a precision
better than $2r_0$. The uncertainty in the total momentum of the $N$ electrons
will be $\hbar/4r_0$, and the uncertainty in the magnetic field generated by
these electrons orbiting on circles of radius $R$ will be of the order of
$e\hbar/(16\pi \epsilon_0c^2m_er_0R^2)$, so that the uncertainty of the 
magnetic flux associated with the shielding electrons over an area $\pi R^2$ 
will be of the order of $\pi\hbar/4e$, which produces a phase uncertainty of 
$\pi/4$, which is sufficient to wipe out the interference pattern
of the incident electron. Thus, it is not possible to ascertain by experimental
observation that the shielding electrons have indeed responded to the field
of an incident electron, and at the same time to preserve the interference
pattern of the incident electrons. This analysis is close to that of Furry
and Ramsey, \cite{14} who have discussed the relation between the AB effect
and Bohr's complementarity principle, and is related to the more general
problem of measurability of fields in quantum mechanics. \cite{15}\\

A similar limitation exists when we try to measure the transient magnetic
field extant in the region between the two paths of an electron interference
experiment in the absence of any shielding. These measurements could be 
conducted for example with the aid of a magnetic semi-string carrying the
flux $F$, which can move freely along the z direction, as shown in Fig. 5.
If a magnetic field $B$ acts for a time $2d/v$ on the test semi-string, the
z momentum transferred to the semi-string is $2\epsilon_0c^2FBd/v$. At
the same time, the flux uncertainty due to the uncertainty in the z position
of the extremity of the semi-string is $F\Delta z/\pi d$, which entails a phase
uncertainty $\Delta \Phi=eF\Delta z/\pi \hbar d$. As it is necessary that 
$2\epsilon_0 c^2FBd/v>\hbar/2\Delta z$, it follows that $B\Delta \Phi>B_e$, 
where $B_e=ev/(4\pi\epsilon_0c^2d^2)$. Thus, a measurement of the magnetic 
field in the region between the interference paths with an accuracy better than
$B_e$ destroys the interference pattern of the incident electron. This
shows that the AB effect is not a vector potential versus magnetic field
case, but rather it is an example of a global phase effect in 
quantum mechanics.\\

\section{Conclusions}

In quantum mechanics, the state of a system composed of several parts is
described by a wave function having a {\it single} phase. Contributions
to this phase arising from the interaction of various parts of the system
may become observable in interference experiments  involving only
a {\it part} of the system. Thus, details of the 
interaction mechanisms which are relevant from the viewpoint of the classical
description, such as a transient magnetic field in a certain region or a
transient change in the velocity of an electron, are not always observable
in quantum mechanics.\\

From this perspective, the Aharonov-Bohm effect appears to be relevant not
so much for the problem of the description of the electromagnetic continuum
by field strengths or electromagnetic potentials, but rather it demonstrates
the global character of the states in quantum mechanics.\\

ACKNOWLEDGMENT\\

This work has been supported by a research grant from the Romanian Academy of
Sciences. The author thanks Dr. Murray Peshkin for a correspondence which
contributed much to the clarification of problems related to the 
subject of this paper.

\newpage
FIGURE CAPTIONS\\

Fig. 1. Electron interference experiment in the presence of a tube of magnetic
flux and of a normal-conducting mesoscopic cylinder.\\

Fig. 2. Incident particle of charge $q$ moving with velocity $v$ along a
straight line which interacts with an enclosed magnetic flux $F$ and with
a ring of uniformly-distributed charge $Q$, rotating with angular velocity
$\Omega$.\\

Fig. 3. Incident particle of charge $q$ and velocity $v$ interacting with
an enclosed magnetic flux $F$, and a particle of charge $Q$ which can move
on a circle of radius $R$. The dashed line shows the second possible path
of the charge $q$ from the incidence region to the observing region.\\

Fig. 4. Phase function $S_N$ for a metallic mesoscopic string at $T$=0 K,
for (a) even values of the number of electrons $N_0$ and (b) odd values of
$N_0$, for magnetic fluxes  $|eF/2\pi\hbar|<1/2$. $S_N$ is a periodic
function of $F$, of period $h/e$.\\

Fig. 5. Mesurement of the transient magnetic field extant in the region
between the arms of an electron interference experiment with the aid of
a magnetic semi-string carrying the flux $F$. The semi-string can move
freely along the z direction, and the uncertainty in the momentum of the
string and the phase difference for the two paths are complementary.\\
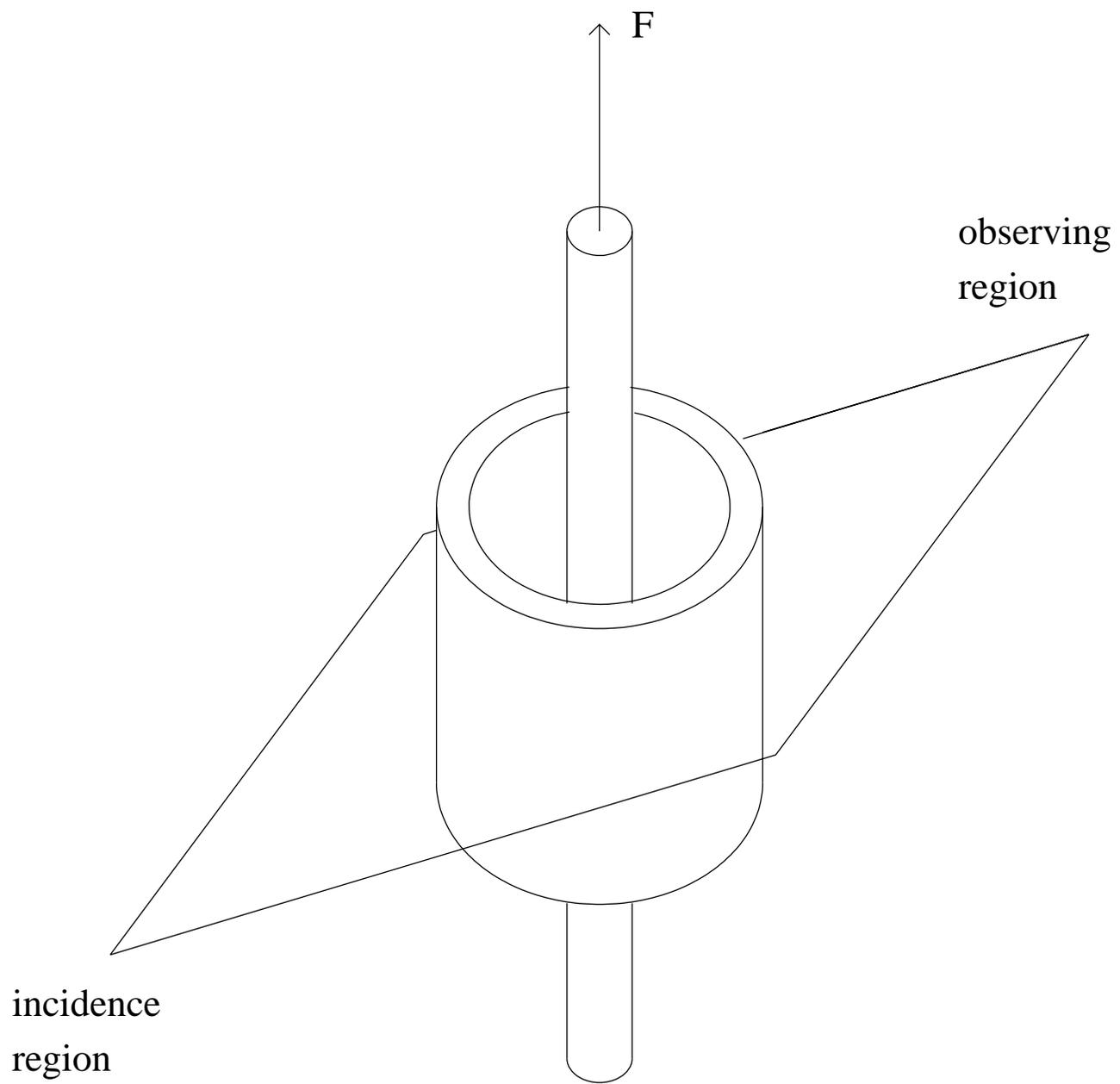
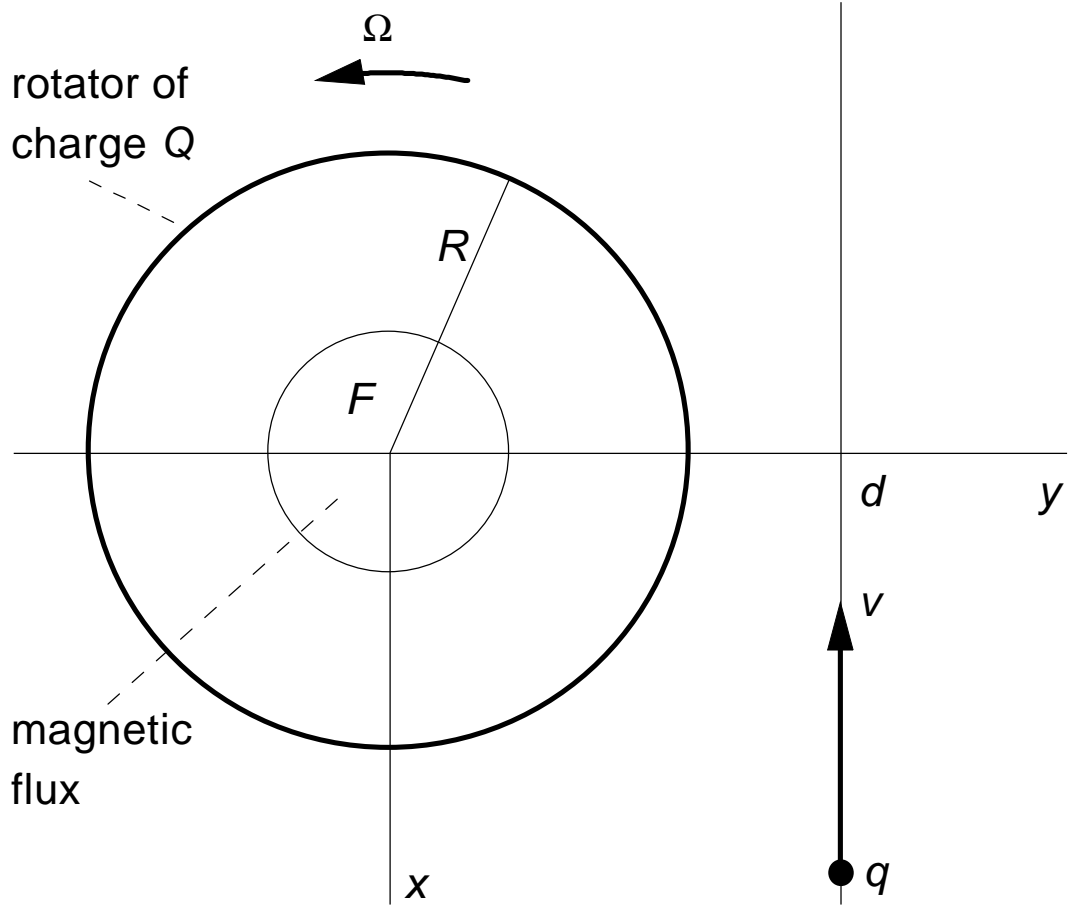
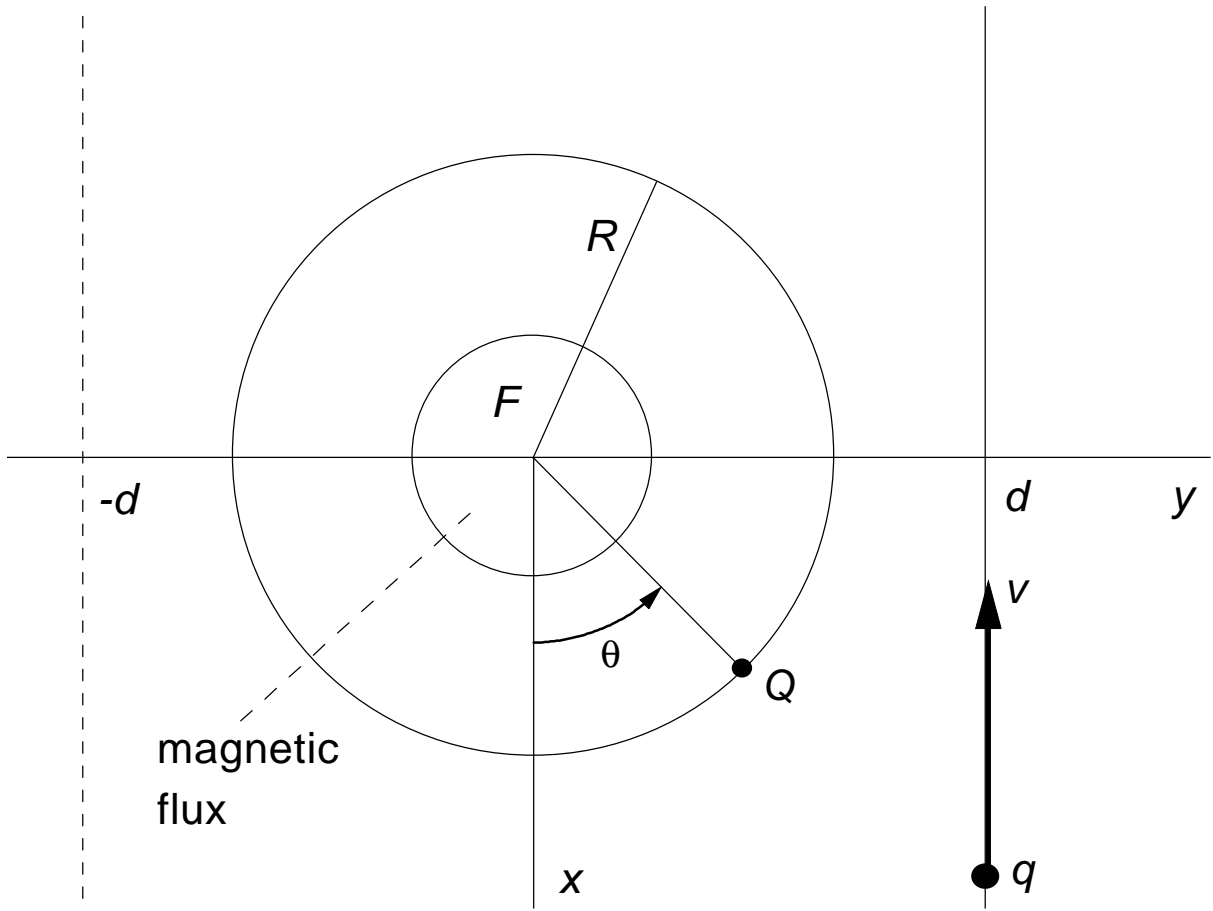
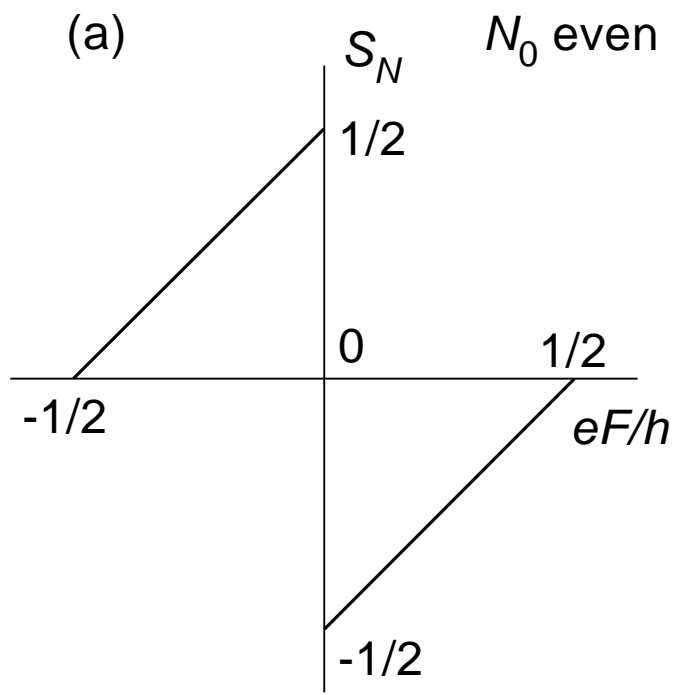
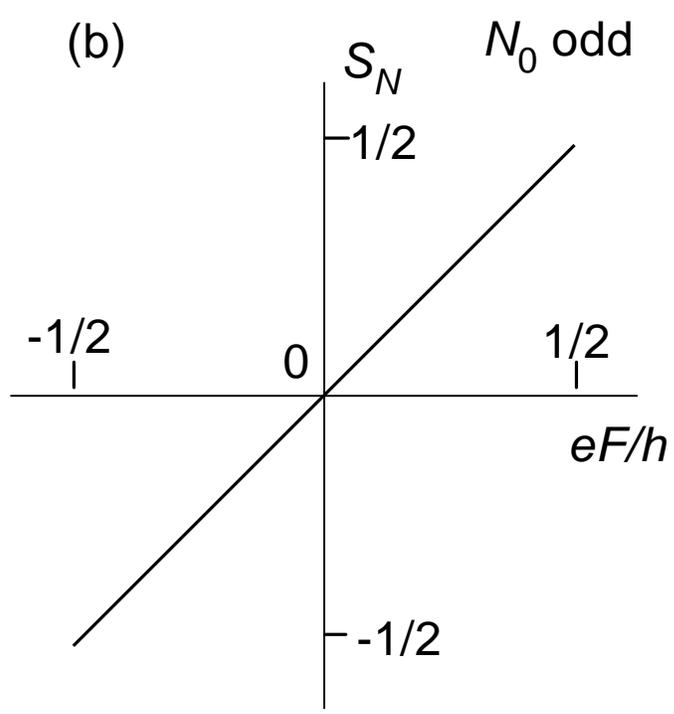
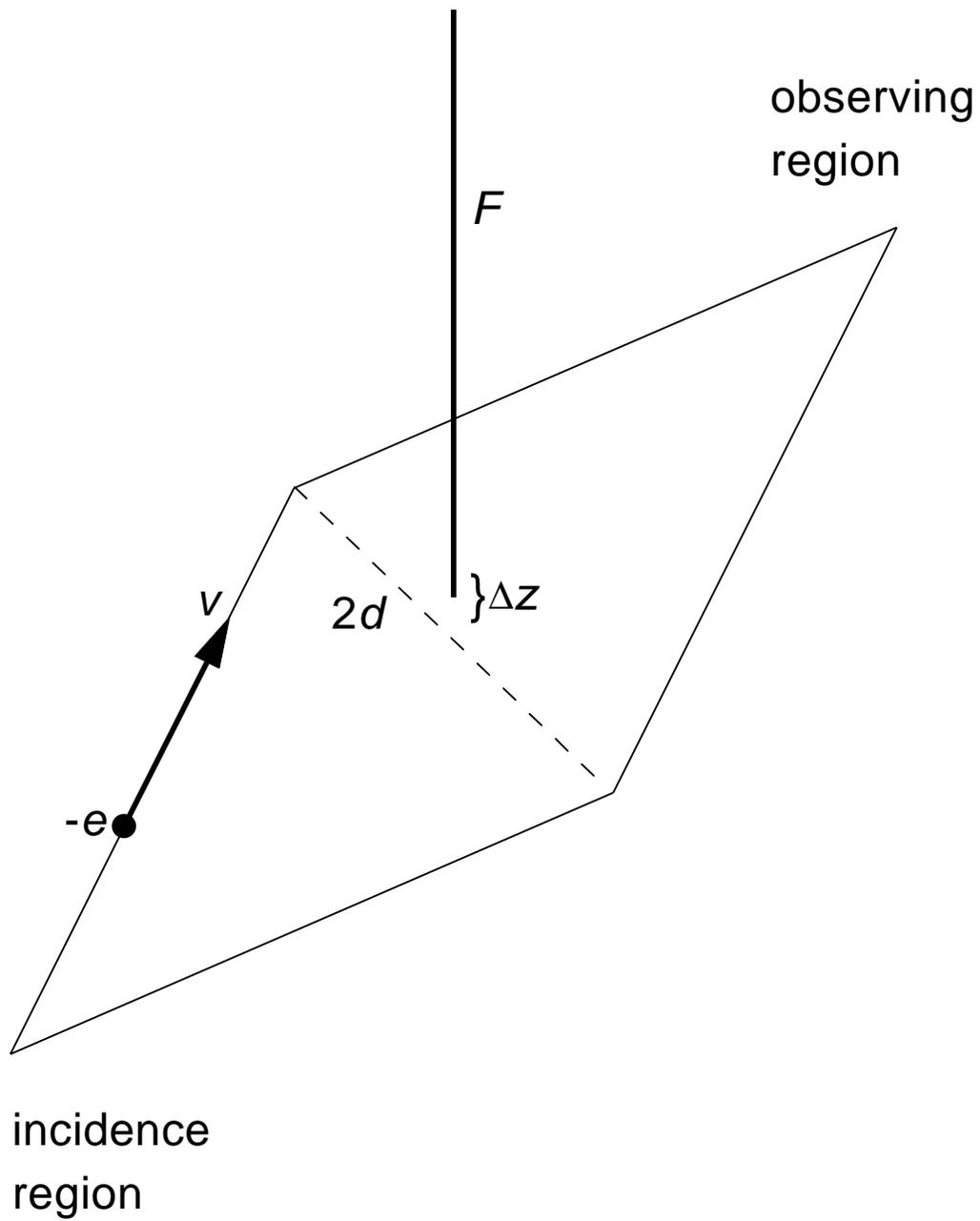
\end{document}